# On Hats and other Covers

## (Extended Summary)

Hendrik W. Lenstra [*]     Gadiel Seroussi[†]

**Abstract.** We study a game puzzle that has enjoyed recent popularity among mathematicians, computer scientist, coding theorists and even the mass press. In the game, $n$ players are fitted with randomly assigned colored hats. Individual players can see their teammates' hat colors, but not their own. Based on this information, and without any further communication, each player must attempt to guess his hat color, or pass. The team wins if there is at least one correct guess, and no incorrect ones. The goal is to devise guessing strategies that maximize the team winning probability. We show that for the case of two hat colors, and for any value of $n$, playing strategies are equivalent to binary covering codes of radius one. This link, in particular with Hamming codes, had been observed for values of $n$ of the form $2^m - 1$. We extend the analysis to games with hats of $q$ colors, $q \geq 2$, where 1-coverings are not sufficient to characterize the best strategies. Instead, we introduce the more appropriate notion of a *strong covering*, and show efficient constructions of these coverings, which achieve winning probabilities approaching unity. Finally, we briefly discuss results on variants of the problem, including arbitrary input distributions, randomized playing strategies, and symmetric strategies.

## 1 Introduction

The following game puzzle has recently received attention from mathematicians, computer scientists, coding theorists, and even from the mass press [1]: A team of $n$ players enters a game room, and each player is fitted with a hat, which is either red or green. A player can see the other players' hat colors, but not his own. Each player is then asked to make a declaration, which must be one of the statements "my hat is red," "my hat is green," or "I pass." All the players must declare simultaneously, and no communication is allowed between them during the game. They are permitted, however, to hold a strategy coordination meeting *before* the game starts. The team wins if at least one player declares his hat color correctly, and no player declares an incorrect color. The distribution of hat color combinations is assumed, for the moment, to be random and uniform, i.e., all $2^n$ combinations are equally likely. The goal of the team is to devise a strategy that maximizes the winning probability.

A winning probability of 50% is guaranteed by a trivial strategy in which a designated player declares "red," and the rest "pass." The puzzle is popularly posed for $n = 3$. In that case, the following strategy yields a 75% winning probability, which turns out to be optimal: Upon observing the hats of the teammates, if a player sees two identical colors, he declares the opposite color, otherwise he passes. Clearly, the team wins whenever the color configuration consists of two hats of the same color and one of the opposite color, and it loses when the three hats are of the same color. Moreover, the following property holds:

> In winning configurations, one player declares the correct color, and the rest pass; in losing configurations all the players declare a wrong color.

[*]Universiteit Leiden, Leiden, The Netherlands, and University of California, Berkeley, U.S.A. e-mail: hwl@math.berkeley.edu

[†]Hewlett-Packard Laboratories, Palo Alto, California, U.S.A. e-mail: seroussi@hpl.hp.com

For any value of $n$, a strategy satisfying this property is said to be *perfect*. In addition to being perfect, the described team strategy for $n = 3$ is also *symmetric*, in that the individual strategies of the players are identical. In general, strategies need not be symmetric; players have identities, and are allowed to use distinct individual strategies. The symmetric constraint for general $n$ is discussed in Section 4.

We will be interested in the behavior of the winning probability for the best strategies as a function of $n$, and in its asymptotic behavior as $n \to \infty$. The rest of the summary is organized as follows: in Section 2 we formalize the notion of a deterministic strategy for the two-color (binary) hats game, and we show that, for any $n$, such a strategy is equivalent to a binary *covering code* [2] of length $n$ and radius one. Consequently, the best strategies are equivalent to the best (smallest) covering codes, and their winning probabilities approach one when $n \to \infty$. The relation between the binary hats game and codes had been previously observed, in particular in relation to Hamming codes; see, e.g. [3, 4], where the game is presented as a "prisoner's puzzle," and optimal strategies are defined for values of $n$ for which Hamming codes exist. In Section 3, we extend the game to $q$ possible hat colors, $q \geq 2$, while leaving the playing rules unchanged. We show that $q$-ary covering codes are not sufficient to describe the best strategies when $q > 2$, and define the notion of a *strong covering*, which obeys much stricter combinatorial constraints. We present efficient constructions of strong coverings of asymptotically vanishing relative size, and whose associated playing strategies, therefore, win with probability approaching one. Finally, in Section 4, we briefly discuss results on other variants of the game, which will be covered in detail in the full version of the paper.

## 2 Optimal Strategies for the Binary Game

Let the symbols 0 and 1 represent the colors in the binary hats game, let $N = \{1, 2, \ldots, n\}$, and let $V_n = \{0, 1\}^n$ denote the $n$-dimensional binary space, $n > 2$.[1] A *deterministic n-player strategy* is a vector $\mathcal{F} = (f_1, f_2, \ldots, f_n)$, where the function $f_i : V_{n-1} \to \{0, 1, pass\}$, $i \in N$, encodes the instructions for player $i$, i.e., upon observing $\mathbf{u} \in V_{n-1}$, player $i$ declares $f_i(\mathbf{u})$. Clearly, given a hats configuration $\mathbf{v} \in V_n$ and a strategy $\mathcal{F}$, it is uniquely determined whether $\mathcal{F}$ wins on $\mathbf{v}$ or not. For a given strategy $\mathcal{F}$, let $W_\mathcal{F} \subseteq V_n$ denote the set of winning configurations for $\mathcal{F}$. Translating the rules of the hats game to this notation, we have

$$\begin{aligned}
(w_1, w_2, \ldots, w_n) \in W_\mathcal{F} &\iff \\
\exists i \in N : f_i(w_1, w_2, \ldots, w_{i-1}, w_{i+1}, \ldots, w_n) &= w_i \\
\wedge \quad \forall j \in N : f_j(w_1, w_2, \ldots, w_{j-1}, w_{j+1}, \ldots, w_n) &\in \{w_j, pass\}.
\end{aligned} \quad (1)$$

A *covering code* of radius $r$, or $r$-*covering* of $V_n$ (in the Hamming metric) is a set $C \subseteq V_n$ such that for all $\mathbf{v} \in V_n$ there is a vector $\mathbf{c} \in C$ satisfying $d(\mathbf{v}, \mathbf{c}) \leq r$, where $d$ denotes the Hamming distance. A comprehensive treatment of covering codes, including an extensive bibliography, can be found in [2]. It follows from the characterization of $W_\mathcal{F}$ in (1) that if $(w_1, w_2, \ldots, w_n)$ is a winning configuration then, for some coordinate $i \in N$, we have $(w_1, w_2, \ldots, \overline{w_i}, \ldots, w_n) \notin W_{\mathcal{F}}$, where $\overline{x}$ denotes the Boolean complement of $x$. Therefore, for any deterministic strategy $\mathcal{F}$, $V_n \setminus W_{\mathcal{F}}$ is a 1-covering of $V_n$, which we shall denote $C_{\mathcal{F}}$. Conversely, let $C$ be a 1-covering of $V_n$, and define the following playing strategy $\mathcal{F}_C$:

> For $i \in N$, and $\mathbf{u} = (w_1, w_2, \ldots, w_{i-1}, w_{i+1}, \ldots, w_n) \in V_{n-1}$, if exactly one value $x \in \{0, 1\}$ satisfies $(w_1, w_2, \ldots, w_{i-1}, x, w_{i+1}, \ldots, w_n) \notin C$, then $f_i(\mathbf{u}) = x$; otherwise, $f_i(\mathbf{u}) = pass$.

---
[1]It is readily verified that for $n \in \{1, 2\}$, the best achievable winning probability is the trivial 50%.

**Proposition 1** *The set of winning configurations of $\mathcal{F}_C$ is $V_n \setminus C$.*

**Proof.** Assume $\mathbf{w} \in V_n \setminus C$. Since $C$ is a 1-covering, there exists an index $i \in N$ such that flipping the $i$th coordinate of $\mathbf{w}$ results in a vector in $C$. Therefore, player $i$ will declare the right color on $\mathbf{w}$. Players $j \neq i$ have at least one way to make the vector fall outside of $C$ (namely, $x = w_j$). Hence, they either guess right or they pass, and $\mathbf{w}$ is a winning configuration. The converse direction is straightforward. □

It follows from the foregoing discussion that there exists a one-to-one correspondence between deterministic strategies and 1-covering codes. Assuming a uniform distribution on the input configurations, the strategy $\mathcal{F}_C$ corresponding to a 1-covering $C$ has *losing probability*

$$P_L = |C|/|V_n| = 2^{-n}|C|. \tag{2}$$

In the sequel, we omit the subscripts $\mathcal{F}$ and $C$ when the correspondence between strategies and coverings is clear from the context.

The strategy for $n = 3$ described in Section 1 corresponds to $C = \{000, 111\}$, i.e., the $[3, 1]$ repetition code, which is also the Hamming code of order 2. This code is invariant under the action of the symmetric group $S_3$ on its coordinates, thus resulting in a symmetric strategy.

Covering codes obey the *sphere-covering* bound, which, for radius one, takes the form

$$|C| \geq 2^n/(n+1). \tag{3}$$

As is well known, the $m$th order binary Hamming code $\mathcal{H}_m$, $m > 1$, has length $n = 2^m - 1$, and it satisfies the sphere-covering bound with equality, namely, it is a *perfect code*. This corresponds to a *perfect strategy* for the hats game, as defined in Section 1. It is also well known that any perfect binary 1-covering must have the same parameters $(n, |C|, d)$ as a Hamming code [2]. Notice that Hamming codes with $n > 3$ are not invariant under $S_n$, and the associated strategies are therefore not symmetric. For values of $n$ for which Hamming codes exist, the strategies attain the optimal losing probability $P_L = (n+1)^{-1}$.

When $n$ is not a Hamming length, we can construct a 1-covering of length $n$ by taking the largest integer $n'$ such that $n > n' = 2^m - 1$, and defining $C = \mathcal{H}_m \times V_{n-n'}$ (a trivial *direct sum* construction [2]). In the corresponding strategy, players 1 through $n'$ play the $n'$-player strategy $\mathcal{F}_{\mathcal{H}_m}$, and ignore teammates $n' + 1$ through $n$. The latter ignore everything, and they always pass. The losing probability for the overall strategy is $P_L = (n' + 1)^{-1}$, which satisfies $(n+1)^{-1} < P_L < 2(n+1)^{-1}$. The factor of 2 on the right hand side of the inequality chain is related to the *density* of the covering, defined as $\mu = |C|(n+1)2^{-n}$. For perfect codes, we have $\mu = 1$, while for the worst values of $n$ in the direct sum construction (those of the form $n = 2^m - 2$), $\mu$ gets arbitrarily close to 2. The latter is true for any linear construction, due to the fact that linear codes only come in sizes that are powers of two. It was shown in [5], however, that one can construct sequences of *non-linear* 1-coverings with $\limsup_{n \to \infty} \mu = 1$. For the derived strategies, the main term of the losing probability $P_L$ approaches, asymptotically, the ideal $(n+1)^{-1}$.

# 3 The $q$-ary Game

## 3.1 Strong coverings

We now consider a game where the hat colors are drawn from a $q$-ary alphabet $Q = \{0, 1, \ldots, q-1\}$, for an arbitrary integer $q \geq 2$. The playing rules remain the same as in the binary case. The trivial strategy where one player always chooses, say, the color 0, and the other players pass, yields a winning probability $1 - P_L = q^{-1}$. Observing how 1-coverings in the binary case led to strategies with winning probabilities approaching one, it might be initially tempting to derive strategies for the $q$-ary game from good $q$-ary 1-coverings. This approach, however, quickly leads to disappointment, even in the case of perfect codes. A closer analysis reveals that a perfect code will lead to situations where, with high probability, one player can guess what color his hat *is not*, while the other players pass. Therefore, strategies derived from $q$-ary 1-coverings approach a winning probability of $(q-1)^{-1}$ (which is of course satisfactory when $q = 2$, but leaves room for improvement otherwise).

As in the binary case, we characterize a deterministic strategy $\mathcal{F}$ in terms of its set $C \subseteq Q^n$ of losing configurations, and its complement $W = Q^n \setminus C$.

**Proposition 2** *Let $\mathbf{w} = (w_1, w_2, \ldots, w_n)$. If $\mathbf{w} \in W$, then there exists a coordinate $i \in N$ such that for all $x \in Q \setminus \{w_i\}$, we have $(w_1, w_2, \ldots, w_{i-1}, x, w_{i+1}, \ldots, w_n) \in C$. Conversely, any pair $C, W$ of complementary sets satisfying this condition defines a valid strategy for the $q$-ary game.* □

We call a set $C$ as characterized in Proposition 2 a *strong covering* of $Q^n$. The following proposition presents an analog of the sphere-covering bound for strong coverings.

**Proposition 3** *Let $C$ be a strong covering of $Q^n$. Then,*

$$|C| \geq \frac{q^n(q-1)}{n+q-1}. \tag{4}$$

Notice that when $q = 2$, the bound (4) coincides with the usual sphere-covering bound, and, in fact, strong coverings coincide in that case with 1-coverings. A strong covering of $Q^n$ is said to be *perfect* if it satisfies (4) with equality.

**Proposition 4** *A strong covering $C$ is perfect if and only if the strategy $\mathcal{F}_C$ is perfect.* □

**Proposition 5** *There are no perfect strong coverings for $q > 2$ and $n > 1$.* □

## 3.2 Winning the $q$-ary game with probability approaching 1

Assume that $Q$ is endowed with an abelian group law. Let $\mathbf{H}_m$ be the parity check matrix of the $m$th order *binary* Hamming code, and let $n = 2^m - 1$. The rows of $\mathbf{H}_m$ are interpreted as characteristic vectors of subsets of $N$, rather than vectors over the finite field $F_2$. $\mathbf{H}_m$ induces a well-defined map $\varphi : Q^n \to Q^m$, with $\varphi(\mathbf{v}) = \mathbf{H}_m \mathbf{v}^T$, where the inner product of a row $[h_{i1}, h_{i2}, \ldots, h_{in}]$ of $\mathbf{H}_m$ and a vector $\mathbf{v} \in Q^n$ is defined as $\sum_{\{j : h_{ij}=1\}} v_j$, with addition taken over the abelian group $Q$. In analogy to the usual linear code terminology, we refer to $\varphi(\mathbf{v})$ as the *syndrome* of $\mathbf{v}$. Clearly, the kernel of $\varphi$ is a subgroup of the $n$-fold direct sum $Q^n$. Assume 0 is the identity of $Q$, and define

$$C = \{\mathbf{v} \in Q^n \;:\; \varphi(\mathbf{v}) \in (Q^*)^m\}, \quad Q^* = Q \setminus \{0\}.$$

**Proposition 6** $C$ is a strong covering of $Q^n$.

**Proof.** Let $\mathbf{w} = (w_1, w_2, \ldots, w_n) \in Q^n \setminus C$. Then, $\varphi(\mathbf{w})$ has some zero entries, and, thus, there is a column $h_j$ of $\mathbf{H}_m$ such that $h_{ij} = 1$ if and only if $\varphi(\mathbf{w})_i = 0$. Therefore, we have $\varphi(\mathbf{w}) + \alpha h_j \in (Q^*)^m$ for all $\alpha \in Q^*$, and, hence, $(w_1, w_2, \ldots, w_{j-1}, x, w_{j+1}, w_n) \in C$ for all $x \in Q \setminus \{w_j\}$. The claim now follows from Proposition 2 and the definition of a strong covering. $\square$

Notice that the "code" $C$ is constructed as the union of those cosets of $\ker(\varphi)$ in $Q^n$ whose corresponding syndromes have only non-zero components.[2] There are $(q-1)^m$ such cosets, out of a total of $q^m$. It follows that the playing strategy corresponding to $C$ has a losing probability

$$P_L = \frac{(q-1)^m}{q^m} = (1 - q^{-1})^{\log_2(n+1)} = (n+1)^{\log_2(1-q^{-1})} \stackrel{n \to \infty}{\longrightarrow} 0. \quad (5)$$

When $q$ is allowed to grow, while still maintaining $q \ll n$, the expression (5) for $P_L$ exhibits an asymptotic behavior of the form $O(n^{-1/q \log 2})$. Both the code $C$ and the associated playing strategy can be efficiently computed.

## 3.3 A generalized construction

Strong coverings with a faster convergence of the relative size (and hence the losing probability) to zero can be obtained by generalizing the construction of Section 3.2. Define

$$C = \{\mathbf{v} \in Q^n \ : \ \varphi(\mathbf{v}) \in (Q^*)^m \ \mathbf{or} \ \mathrm{wt}(\varphi(\mathbf{v})) < \beta m\}, \quad 0 \leq \beta < 1. \quad (6)$$

Here, $\beta$ is a parameter to be optimized, and $\mathrm{wt}(\cdot)$ denotes the Hamming weight. The construction of Section 3.2 corresponds to $\beta = 0$. At first sight, using a value of $\beta > 0$ seems counter-intuitive, as it appears to increase the size of $C$. However, the inclusion of low-weight syndromes in the definition of $C$ allows us to use a subset of the columns of $\mathbf{H}_m$, since the high-weight columns are no longer needed to match low-weight syndromes produced by vectors in $Q^n \setminus C$. This can reduce the value of $n$ for a given order $m$, which in turn might improve the vanishing rate of $|C|/q^n$.

**Proposition 7** Let $q > 5$ and let $\gamma_q$ be the nonzero root of the equation $h_2(\gamma) = \gamma \log_2(q-1)$, where $h_2(\cdot)$ is the binary entropy function. Set $\beta = 1 - \gamma_q$, and redefine $\varphi$ in (6) as $\varphi(\mathbf{v}) = \hat{\mathbf{H}}_m \mathbf{v}^T$, where $\hat{\mathbf{H}}_m$ is formed from the subset of columns $h_j$ of $\mathbf{H}_m$ such that $0 < \mathrm{wt}(h_j) \leq \gamma_q m$. Then, $C$ is a strong covering of $Q^n$ satisfying $|C|q^{-n} = O(n^{-c})$, with

$$c \geq \left(1 - \frac{1}{q}\right)\left(1 + \frac{1}{2(q-1)}\right)\frac{1}{e\log(q-1)}. \quad (7)$$

$\square$

For values of $q < 11$, it is advantageous to choose $\beta = 0$ and revert to the construction of Section 3.2. However, for $0 \ll q \ll n$, the dominant term in the exponent $c$ of (7) is of the form $1/(e\log(q-1))$, which compares favorably with $1/(q\log 2)$ in the first construction.

Noga Alon [6], using a random-coding argument, showed the existence of strong coverings with

$$P_L \leq \frac{(q-1)\log n + 1}{n} + (1 - q^{-1})^n. \quad (8)$$

For fixed $q$, the main term of this upper bound behaves like $(q-1)\log n/n$, compared to $(q-1)/n$ for the sphere-covering-like lower bound (4). Finding an efficient construction that attains the bound (8), and closing the gap between (4) and (8) remain open problems.

---

[2] This runs somewhat contrary to the usual construction of linear codes. In particular, for $q = 2$, the 1-covering produced by the construction is the coset of the Hamming code whose syndrome is all-ones.

# 4  Variations

We briefly describe some results on variations of the game described in Section 1. These results will be covered in detail in the full version of the paper.

**Non-uniform input distributions.** The playing strategies described in previous sections are easily modified to accommodate non-uniform distributions of the hat color configuration. We assume the distribution used is unknown to the players. However, all they need to do is choose, in their strategy session, a random translation to be applied to $C$ at game time. The translation cancels the effect of any non-uniformity in the input distribution, without affecting the covering properties of the original code.

**Randomized playing strategies.** In this setting, players choose their declarations at random, conditioned on their observations. It is proved that from the perspective of each player, the overall probability of team success is a linear function of the parameters of the random strategy, which are in turn subject to linear constraints. It follows that an optimal winning probability is attained at a vertex of the convex polyhedron defined by these constraints, i.e., for a *deterministic* playing strategy. It follows that randomization does not help the team.

**Symmetric strategies.** Players are indistinguishable, and are forced to use identical strategies, based on *counts* of observed colors. It is proved that for the binary game, the best symmetric strategy achieves a winning probability approaching $2/3$ as $n \to \infty$. This implies that the best unconstrained strategies for large values of $n$ must be asymmetric. In fact, $n = 3$ is the largest value of $n$ for which a symmetric strategy is optimal.

**Zero-information strategies.** Players make a random declaration, without access to any information. A winning probability of $1/4$ is asymptotically attainable and optimal.

**Acknowledgments.** Thanks to Joe Buhler for introducing us to the problem and for many stimulating discussions, and to Noga Alon for his communication [6].

# References

[1] S. ROBINSON, "Why mathematicians now care about their hat color," The New York Times, Science Times section, p. D5, April 10, 2001.

[2] G. COHEN, I. HONKALA, S. LYTSIN AND A. LOBSTEIN, *Covering Codes*, North-Holland, Amsterdam, 1997.

[3] T. EBERT, *Applications of Recursive Operators to Randomness and Complexity*, Ph.D. Thesis, University of California at Santa Barbara, 1998.

[4] T. EBERT AND H. VOLLMER, "On the autoreducibility of random sequences," *Proc. 25th International Symposium on Mathematical Foundations of Computer Science*, Springer Lecture Notes in Computer Science Vol. 1893, 333-342, 2000.

[5] G.A. KABATYANSKII AND V.I. PANCHENKO, "Unit sphere packings and coverings of the Hamming space," *Probl. Peredachi Informatsii*, vol. 24, No. 4, pp. 3–16, 1988. Translated in *Probl. Inform. Transm.*, vol. 24, No. 4, pp. 261–272.

[6] N. ALON, "A comment on generalized covers," e-mail communication, June 2001.